RESEARCH

# Human-Like Attention? A Psychophysical Comparison of Visual Search in Humans and MLLMs

R. Zhang[1] · J. C. F. de Winter[1] · D. Dodou[1] · H. Seyffert[1] · Y. B. Eisma[1]

Received: 23 December 2025 / Accepted: 20 July 2026

**Abstract**
Visual search is a fundamental cognitive ability. This study investigates whether Multimodal Large Language Models (MLLMs) exhibit human-like difficulty signatures in visual search tasks. We compared search performance of humans ($n$ = 1,250) and MLLMs using identical 2D and 3D stimuli across different set sizes. Both groups showed efficient performance in feature searches, most clearly when the target had a unique color, but performance degradation in conjunction searches as set sizes increased. Additionally, we found strong correlations between human and MLLM error rates ($\rho$ = 0.82), which suggests that MLLMs are sensitive to similar objective complexities, such as stimulus heterogeneity. However, differences were found as well: whereas humans invested extra search time to respond accurately on target-absent trials, MLLMs exhibited extreme present/absent response biases in complex searches. We conclude that MLLMs replicate high-level human performance signatures, yet their underlying computations differ significantly.

## Introduction

Extracting relevant information from visual scenes is fundamental to intelligent perception (Krizhevsky et al., 2012; Marr, 2010). In cognitive psychology, visual search paradigms have been used to reveal limitations regarding how humans encode features, select targets, and bind them into objects (Duncan & Humphreys, 1989; Treisman & Gelade, 1980; Wolfe, 2021).

Classic visual search theories distinguish between efficient feature search and effortful conjunction search. In feature search, a unique target often pops out rapidly, yielding shallow set-size effects (i.e., near-flat RT slopes under high discriminability conditions) (Treisman & Gelade, 1980). Conjunction search, by contrast, requires binding of multiple features. Target-absent trials often show steeper set-size costs than target-present trials (commonly approximating a ~ 2× slope ratio), a pattern historically attributed to self-terminating serial scanning (Treisman & Gelade, 1980; Wolfe et al., 1989), though capacity-limited parallel models offer an alternative explanation (McElree & Carrasco, 1999; Thornton & Gilden, 2007).

Contemporary models, such as Guided Search 6.0 (Wolfe, 2021), propose that visual search is guided by a "priority map" that is shaped by bottom-up salience, top-down feature guidance, scene structure, search history, and relative item value (Wolfe & Horowitz, 2017). In terms of feature guidance, basic features, such as color and orientation, offer strong signals (Wolfe & Horowitz, 2017; Wolfe et al., 1992), while others such as material provide weaker input and often yield inefficient search (Wolfe, 2021; Wolfe & Myers, 2010). This guidance is constrained by target-distractor relationships, where search performance is affected by similarity and heterogeneity (Calder-Travis & Ma, 2020; Duncan & Humphreys, 1989; Lleras et al., 2019), and improves when targets are linearly separable from distractors (Bauer et al., 1996; D'Zmura, 1991; but see Xu et al., 2021). Furthermore, conjunctions within a single dimension (e.g., color × color) are less efficient than those across different dimensions (e.g., color × orientation) (Wolfe et al., 1990). Finally, consistent with the role of scene structure, 3D depth cues and geometry can facilitate item segregation, but can also hinder performance through occlusion (Enns & Rensink, 1990; Plewan & Rinkenauer, 2021), with additional effects arising from texture and stereopsis (Aks & Enns, 1996; Finlayson & Grove, 2015; O'Toole & Walker, 1997; Previc & Blume, 1993).

✉ J. C. F. de Winter
j.c.f.dewinter@tudelft.nl

[1] Faculty of Mechanical Engineering, Delft University of Technology, Delft, The Netherlands

Recent developments in artificial intelligence have led to the development of Multimodal Large Language Models (MLLMs), which extend traditional LLMs by incorporating visual inputs (Alayrac et al., 2022; Liu et al., 2023). This integration enables the execution of tasks that require joint vision-language understanding (Bordes et al., 2024; Yin et al., 2024). Because MLLMs are trained on large human-generated datasets, they can exhibit behavioral signatures that resemble human performance on specific tasks (Campbell et al., 2024; Yiu et al., 2025). However, it remains unclear whether these similarities arise from shared cognitive mechanisms or simply reflect the objective complexity of visual signals that would constrain any ideal observer (cf. Geisler, 2011). Resolving this question could reveal a connection between statistical learning and biological vision, or instead expose architectural limitations of MLLMs. Understanding these similarities and differences can inform efforts to align MLLMs more closely with human cognition, which is vital for real-world applications such as robotics and assistive technology.

Research suggests that humans and MLLMs exhibit overlapping limitations, raising the possibility of shared information-processing bottlenecks (Burden et al., 2025). Campbell et al. (2024) found human-like set-size costs in conjunction search, whereas Yiu et al. (2025) found that while the strongest model tested, o1, correlates with human performance on simple surface-level attributes such as color and size, it underperformed in tasks that require deeper cognitive processing, such as mental rotation and counting. Similarly, studies report deficits in binding, occlusion, and cognitive conflict resolution (Luo et al., 2025; Tangtartharakul & Storrs, 2026).

Human visual search is constrained by non-uniform spatial resolution in peripheral vision and the need to reposition high-acuity foveal vision via saccades (Rosenholtz, 2016). This does not need to imply that eye movements themselves are the bottleneck: covert attentional selection can determine the effective units of selection (e.g., groups vs. items) even when displays are dense (Treisman, 1982). By contrast, most current MLLMs encode a fixed-resolution image into discrete patch tokens (or tiled variants) and process them in a feed-forward manner, rather than through gaze-contingent sampling (Dosovitskiy et al., 2021). Architectural analyses therefore caution against equating transformer self-attention with human visual attention; in vision transformers it often behaves more like similarity-based token interaction/perceptual grouping than a goal-directed attentional controller (Mehrani & Tsotsos, 2023; Zhao et al., 2024). Patch-based encoding can also reduce available fine spatial detail (Gemma Team, 2025), consistent with documented failures on small-object and basic visual-cognition tasks (Huang et al., 2025; Rahmanzadehgervi et al., 2024; Zhang et al., 2024).

Consistent with these differences in sensory and computational limitations, evaluations report that MLLMs have particular difficulty with tasks that require integrating multiple objects and attributes (especially under occlusion, high item counts, or relational binding demands) where humans can often exploit perceptual organization to reduce the effective search space (Huang et al., 2025; Li, 2023; Treisman, 1982). More broadly, tests of deep neural networks show mixed evidence for human-like Gestalt grouping: some architectures exhibit sensitivity to proximity/linearity, but such effects may emerge primarily at late decision stages rather than as an intermediate representation that organizes scenes for subsequent selection and recognition (Biscione & Bowers, 2023).

Inspired by cognitive benchmarks like VISFACTOR (Huang et al., 2025), we developed a paradigm to evaluate visual search in humans and MLLMs using synthetic 2D and 3D scenes. The task posed a binary question: "Is there a solid yellow square (2D) or cube (3D)?" We tested nine conditions spanning feature and conjunction searches across varying set sizes (4–256 items) and balanced target presence (50% present/absent). 3D scenes incorporated realistic depth cues, including occlusion, shadows, and perspective. We hypothesized that MLLM accuracy profiles would resemble human patterns: near-perfect performance in feature search, but declining accuracy with increasing set size in conjunction search. Furthermore, we predicted that 3D scenes would pose greater challenges for both groups. Specifically, the processing demands introduced by depth cues and occlusion were expected to lower search efficiency, leading to higher RTs and error rates, compared to 2D counterparts.

Although iterative agents can simulate serial search (Wu & Xie, 2024), and visual scaffolds (Izadi et al., 2025) or chain-of-thought reasoning (Gao et al., 2025; Zhang et al., 2025) can aid attention, these approaches may confound the analysis of inherent architectural limitations. Moreover, even with chain-of-thought prompting, models often fail to align entities across modalities in complex scenes (Alonso et al., 2025). Therefore, in the present study, we probe the baseline capabilities of MLLMs without augmentation to assess whether they guide search via human-like feature relationships (cf. Becker et al., 2014; D'Zmura, 1991; Sobel & Cave, 2002) or rely on superficial patterns that break under complexity (Rahmanzadehgervi et al., 2024).

This work extends prior research in four ways. First, by using identical stimuli for both groups, we enable a comparison of search efficiency and target-absent costs, extending recent findings on MLLM visual deficits (Burden et al., 2025; Campbell et al., 2024; Tangtartharakul & Storrs, 2026). Whereas Campbell et al. (2024) collected no human visual-search data and Burden et al. (2025) compared models with a small human sample on a stimulus subset, scored on accuracy alone in three target-localization tasks rather than present/absent detection, we collect a large human sample ($n$=1,250) on identical 2D and 3D stimuli. This yields correlations of MLLM error rates with human error rates and response times, and separate target-present/absent analyses that reveal a non-human-like present/absent reversal. Second, we manipulated distractor heterogeneity and 3D structure across a wide set-size range (4–256) to isolate binding interference from depth factors. Third, we established a new benchmark investigating fundamental attributes (color, shape, material) across MLLMs. Finally, we did not use external zoom-and-crop pipelines (e.g., Wu & Xie, 2024) in order to isolate the baseline capacity limits of MLLMs, allowing us to test whether human-like binding patterns generalize across 2D and 3D contexts.

## Methods

### Experimental Conditions

We designed nine conditions (C1–C9) spanning efficient feature search and effortful conjunction search, using a fixed target of a yellow square (2D) or cube (3D) (see Table 1; Fig. 1).

In the feature conditions (C1–C6), the target served as a singleton on at least one dimension (e.g., color or shape), allowing for parallel preattentive guidance. In the conjunction conditions (C7–C9), no target item was unique on any single dimension. Successful performance therefore required binding features (e.g., color+shape). We systematically varied distractor heterogeneity, predicting that increased heterogeneity (e.g., C8, C9) would degrade performance.

Each condition was tested across seven set sizes (4, 8, 16, 32, 64, 128, 256) and balanced target presence (50% present/absent). We generated 25 versions of every configuration by randomizing object positions, yielding 6,300 unique images (9 conditions × 2 target presence levels × 7 set sizes × 2 dimensionalities × 25 versions). Separate 2D and 3D datasets were created, with the 3D set rendered in Blender (Blender Foundation, 2017) based on the CLEVR framework (Johnson et al., 2017; see also Campbell et al., 2024) to incorporate depth cues.

### MLLM Evaluation

We evaluated three models: gemini-2.0-flash-lite, gemini-2.5-flash-lite (with thinking budget=0), and gemma-3-27b-it, accessed in August 2025 (Google, 2025a, b; Google DeepMind, 2025). To minimize stochasticity, each image was queried 10 times at temperature 0. These lightweight, widely used instruction-tuned models were selected as representative, cost-accessible MLLMs evaluated under identical conditions. We appended one to ten "." symbols to the prompt, creating ten deterministic prompt variants under temperature-0 decoding. This technique exploits the sensitivity of LLMs to minor perturbations, allowing us to generate a distribution of responses under deterministic decoding conditions (Salinas & Morstatter, 2024; Sclar et al., 2024). By aggregating these predictions, we established a performance estimate, akin to the self-consistency prompting principle (Wang et al., 2023).

**Table 1** Experimental conditions, including example composition with target present and set size=128

| Condition | Label | Distractor composition | Search type | Predicted search efficiency |
|---|---|---|---|---|
| C1 | F_ColorShape | 127 solid blue circles (spheres) | Redundant feature search. Target is unique in both color and shape. | Easy |
| C2 | F_Color | 127 solid blue squares (cubes) | The base version of a color feature search. | Easy |
| C3 | F_Shape | 127 solid yellow circles (spheres) | The base version of a shape feature search. | Medium |
| C4 | F_Material | 127 textured yellow squares (cubes) | The base version of a material feature search. | Medium |
| C5 | F_Color_Het | 64 solid blue squares (cubes) & 63 solid blue circles (spheres) | Color feature search with heterogeneous distractor shapes. | Easy |
| C6 | F_Shape_Het | 64 solid yellow circles (spheres) & 63 solid blue circles (spheres) | Shape feature search with heterogeneous distractor colors. | Medium |
| C7 | C_ColorShape | 64 solid yellow circles (spheres) & 63 solid blue squares (cubes) | The base version of a two-feature conjunction search. | Hard |
| C8 | C_ColorShape_Het | 43 solid blue squares (cubes), 42 solid yellow circles (spheres) & 42 solid blue circles (spheres) | The same as C7, but with added heterogeneous distractors (blue circles). | Hard |
| C9 | C_ColorShapeMaterial | 43 solid blue squares (cubes), 42 solid yellow circles (spheres) & 42 textured yellow squares (cubes) | A three-feature (triple) conjunction search. | Very hard |

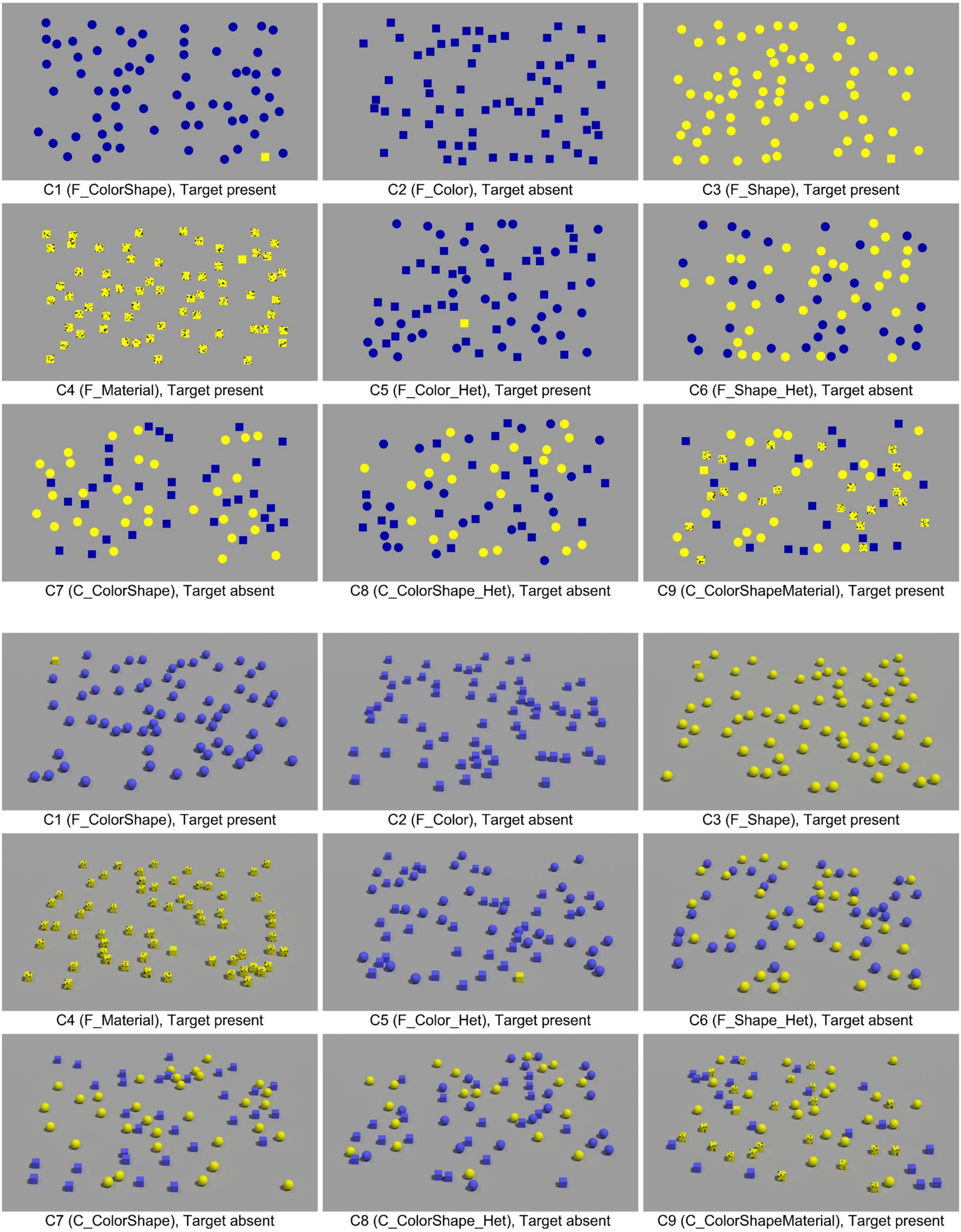

**Fig. 1** Images used in the practice block covering all 9 conditions, in a 2D layout (top 9 figures) or 3D layout (bottom 9 figures)

We used the following prompt to instruct the models:

*In this task, you will search for a solid yellow {block} without any texture on its face in an image.*

*{variations}.*

*Solid yellow {block}: Output 0 for absent or 1 for present.*

The placeholder {block} was adapted to the dimension (square/cube). Responses were recorded per image repetition to assess error rates; of 188,998 recorded calls, 54 failed (0.03%) and were excluded.

### Human Evaluation

We recruited 1,250 participants via the Prolific academic research platform (Douglas et al., 2023; Peer et al., 2022). Participants were required to reside in one of the following English-speaking countries to be eligible: Australia, Canada, Ireland, New Zealand, the United Kingdom, or the United States. The study launched on September 15, 2025, with the first participant starting at 19:17 CEST and the last participant starting at 20:51 CEST. Participants were compensated with £2.10 for their contribution. This study received ethical approval from the TU Delft Human Research Ethics Committee (Reference Number: 4182), and informed consent was obtained from all respondents through a dedicated item at the beginning of the questionnaire.

Participants were randomly assigned to either the 2D or 3D branch. They first completed a mixed-condition practice block to familiarize them with the stimuli. The practice block consisted of nine trials at set size 64 (covering all conditions, with a mix of present/absent), see Fig. 1. This was followed by nine randomized experimental blocks (one per condition). Each block contained 14 trials (7 set sizes × 2 target presence levels). Participants were randomly assigned one of 25 image versions.

Trials began with a 1000 ms fixation cross, followed by the search display. Participants indicated target presence or absence via keypress ('A' for 'absent', 'P' for 'present'), and received immediate textual feedback ("Correct"/"Incorrect").

The sample consisted of 50.6% female, 48.2% male, and 1.2% identified as non-binary or preferred not to disclose their gender. The participants' mean age was 44.28 years (Median = 43, *SD* = 13.34), and they completed the experiment in an average of 14.10 min (Median = 12.22, *SD* = 6.27). Most participants (62.2%) completed the study on a laptop, while 30.3% used a desktop computer. A minority used a laptop connected to an external monitor or keyboard (7.2%) or a tablet (0.3%).

### Analysis

All practice trials were excluded. We excluded trials with RTs < 150 ms or > 20,000 ms using a single, shared inclusion mask that was applied across all RT and error-rate analyses. Out of 157,500 experimental trials, 241 were unavailable due to an unspecified error and 217 were excluded by the RT bounds, resulting in a final dataset of 157,042 trials.

We visualized human performance via mean RT and error rates as functions of set size. Mean RTs were computed as arithmetic means across all included trials, i.e., regardless of response accuracy, so that the same trials underlie the RT and error-rate analyses. To assess search efficiency limits, we analyzed RT and error rates for target-absent vs. target-present trials at the largest set size (256).

MLLM performance was visualized using error rates averaged across target presence (present/absent) within each model and then averaged across the three models. This averaging approach was chosen to provide a more robust measure of the capabilities of the MLLMs, minimizing the influence of model-specific quirks (for similar approaches, see Li et al., 2024; Lu et al., 2024).

Finally, we calculated Spearman rank correlations between human and MLLM performance across the 18 cells (2 dimensionalities × 9 conditions), with each cell's error rates and mean RT pooled across set sizes, to quantify the relationship between human and model error rates, and between human RT and model error rates.

## Results

### Human Performance

Search functions (Figs. 2 and 3) revealed distinct efficiency patterns. Specifically, conditions C1, C2, and C5 exhibited near-zero slopes, consistent with parallel feature search where the unique color guides attention efficiently. In contrast, all other conditions showed substantial set-size effects, with systematic increases in RT and error rates consistent with inefficient (capacity-limited) search. Difficulty ranged from low in simple shape searches (C3, C6), to moderate in material (C4) and shape-color conjunctions (C7, C8), to highest in the triple conjunction search (C9). For shape-based searches (C3, C6), error rates were higher in 3D than in 2D, possibly because occlusion, foreshortening, and shading reduce the discriminability of the sphere/cube silhouettes.

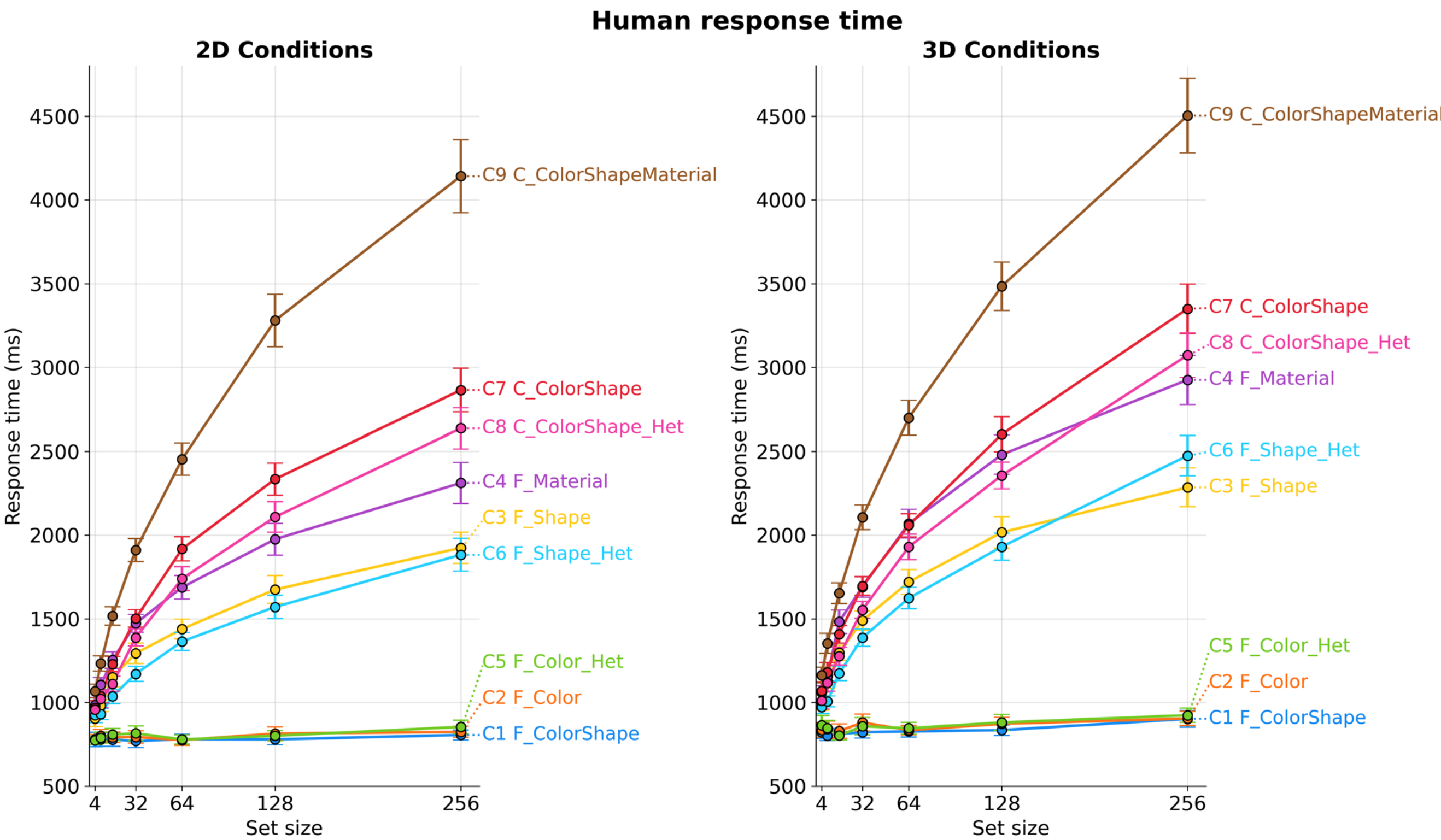

**Fig. 2** Mean human response time as a function of set size for the different experimental conditions. Left: 2D conditions, Right: 3D conditions. Error bars are participant-level 95% confidence intervals

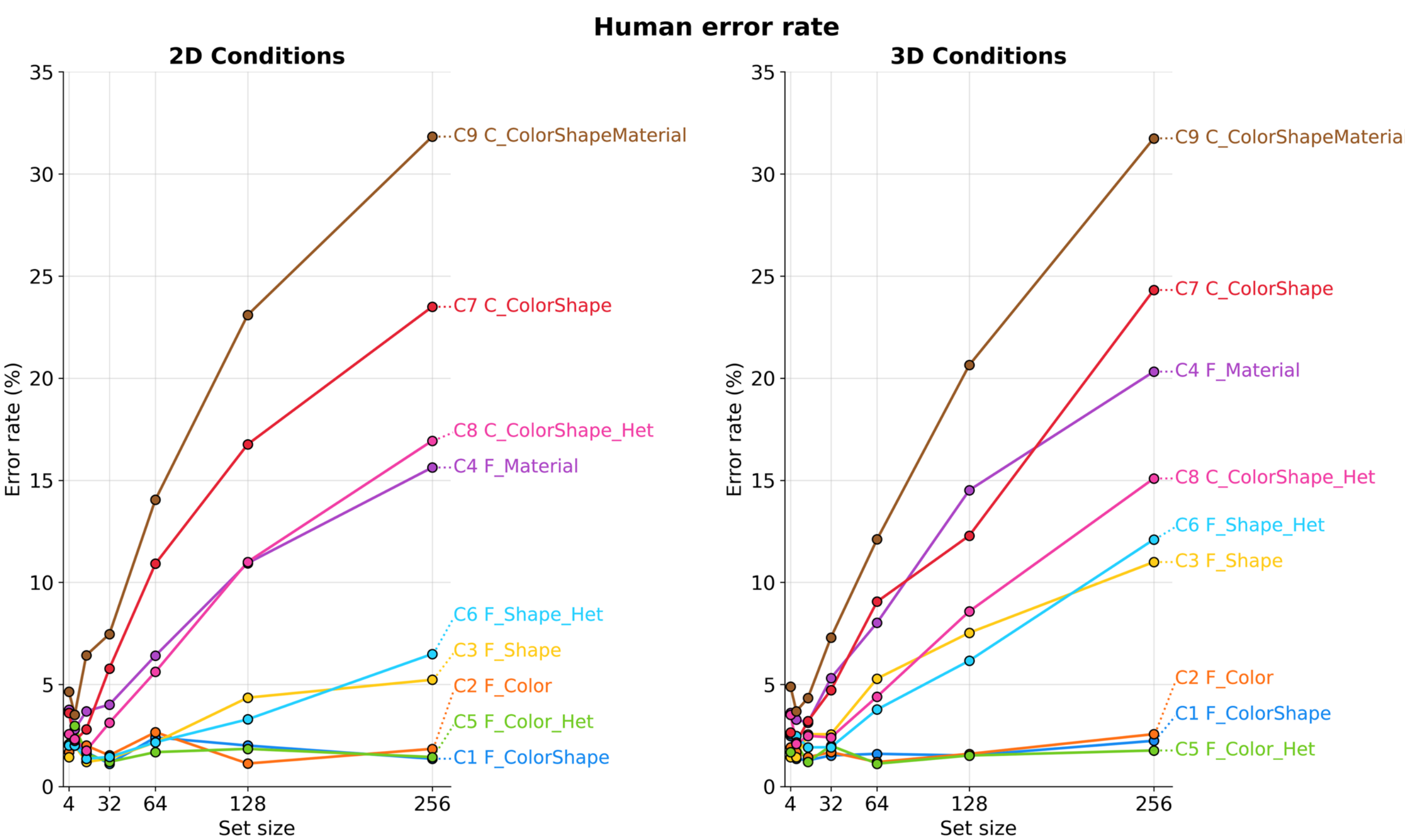

**Fig. 3** Mean human error rate as a function of set size for the different experimental conditions. Left: 2D conditions, Right: 3D conditions

As shown in Fig. 4, target-absent trials were consistently more accurate and, in all but the easiest searches, slower than target-present trials, where participants frequently terminated search prematurely. Generally, 3D conditions elicited longer RTs and higher error rates than their 2D counterparts, confirming the added cost of processing depth and occlusion.

## MLLM Performance

MLLM error rates strongly paralleled human difficulty hierarchies (Fig. 5). Feature searches defined by a unique color (C1, C2, C5) yielded very low error rates (<5%) across all set sizes. Shape-based feature searches (C3, C6) were also highly accurate, with a moderate increase in errors at larger set sizes in the 3D condition. Material search (C4) proved considerably harder.

Conjunction conditions (C7–C9) showed performance deterioration, with error rates reaching 40–50% at large set sizes in heterogeneous conditions (C8, C9), indicating a failure to bind multiple features efficiently in complex scenes. Unlike humans, MLLMs found C8 more difficult than C7. The introduction of a third distractor type in C8 led to an increase in error rates, suggesting that MLLMs do not benefit from feature-based grouping to reduce the effective set size.

Error patterns regarding target presence diverged from human behavior (Fig. 6). Averaged across models, simple conjunctions (C7) showed a standard target-absent cost: the MLLMs were highly accurate when the target was present (≤9% errors) but struggled to confirm its absence, with error rates exceeding 50% at the largest set size. However, complex conditions revealed non-human-like biases. In the 3D triple-conjunction (C9), the MLLMs frequently missed targets (high error rate) but achieved near-perfect accuracy when the target was absent, whereas in C8, they defaulted to "present" responses instead. An extreme bias was not confined to conjunction searches: in material search (C4), the models almost never false-alarmed on target-absent trials but missed the target on about 50% of target-present trials at set size 256, effectively defaulting to "absent". This suggests MLLMs may not employ the same strategic "quitting threshold" as humans, instead showing extreme response biases. The impact of 3D dimensionality was mixed, increasing errors in feature tasks but showing context-dependent effects in conjunctions.

## MLLMs vs. Humans

We found strong positive correlations between human and MLLM performance: ρ=0.91 for human RT vs. MLLM error, and ρ=0.82 for human error vs. MLLM error ($n$=18 cells, pooled across set sizes). This indicates that the objective complexity factors driving human difficulty, such as stimulus similarity and heterogeneity, constrain MLLMs in a similar manner. Figure 7 depicts these relationships at the largest set size, where between-condition differences are most pronounced.

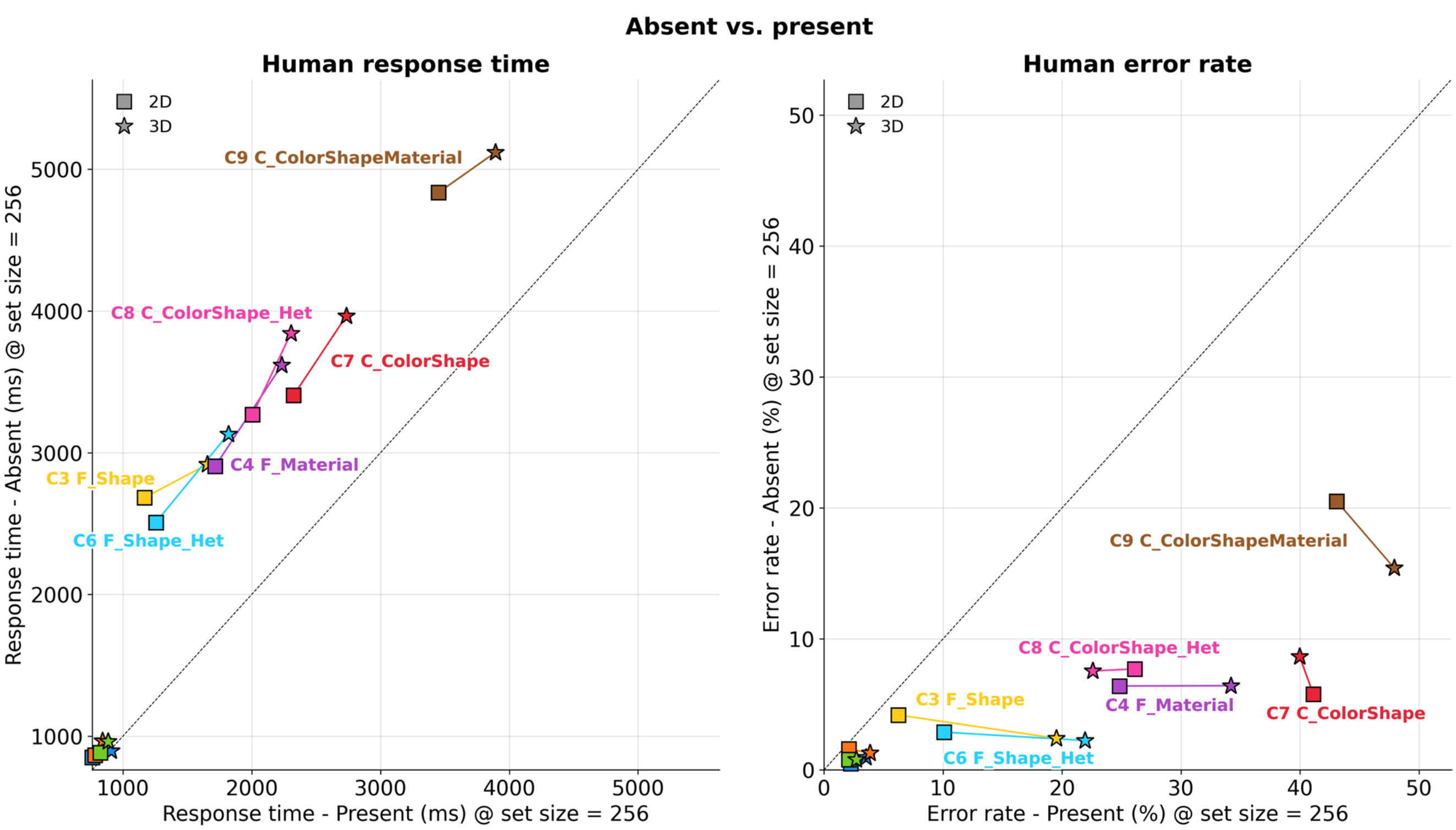

**Fig. 4** Mean human response time and human error rate for target-absent trials versus target-present trials, at the largest set size. Corresponding 2D and 3D conditions are connected by a line. The diagonal line is the line of unity

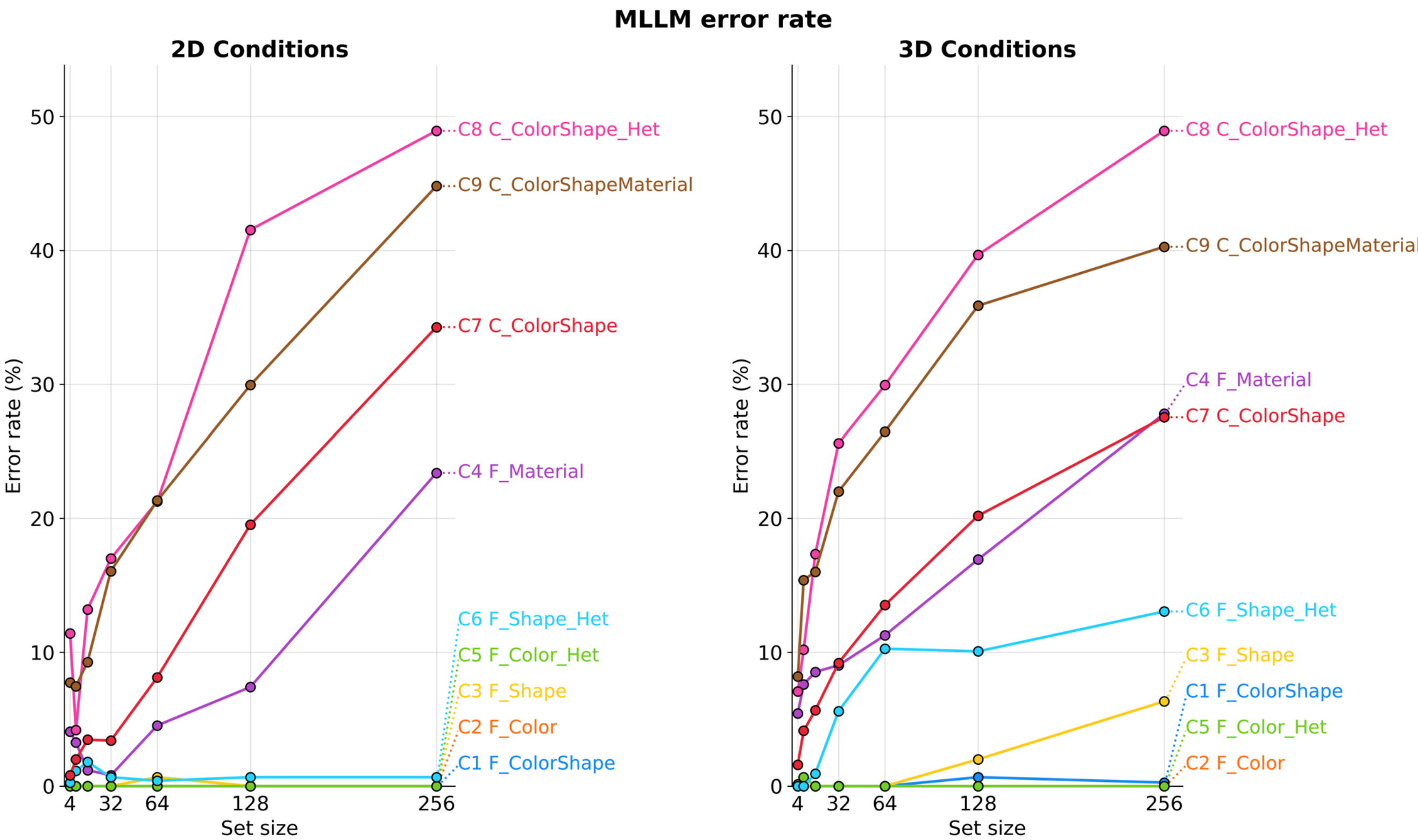

**Fig. 5** MLLM error rate as a function of set size for the different experimental conditions. Left: 2D conditions, Right: 3D conditions

**Fig. 6** MLLM error rate for target-absent trials versus target-present trials, at the largest set size. Corresponding 2D and 3D conditions are connected by a line. The diagonal line is the line of unity

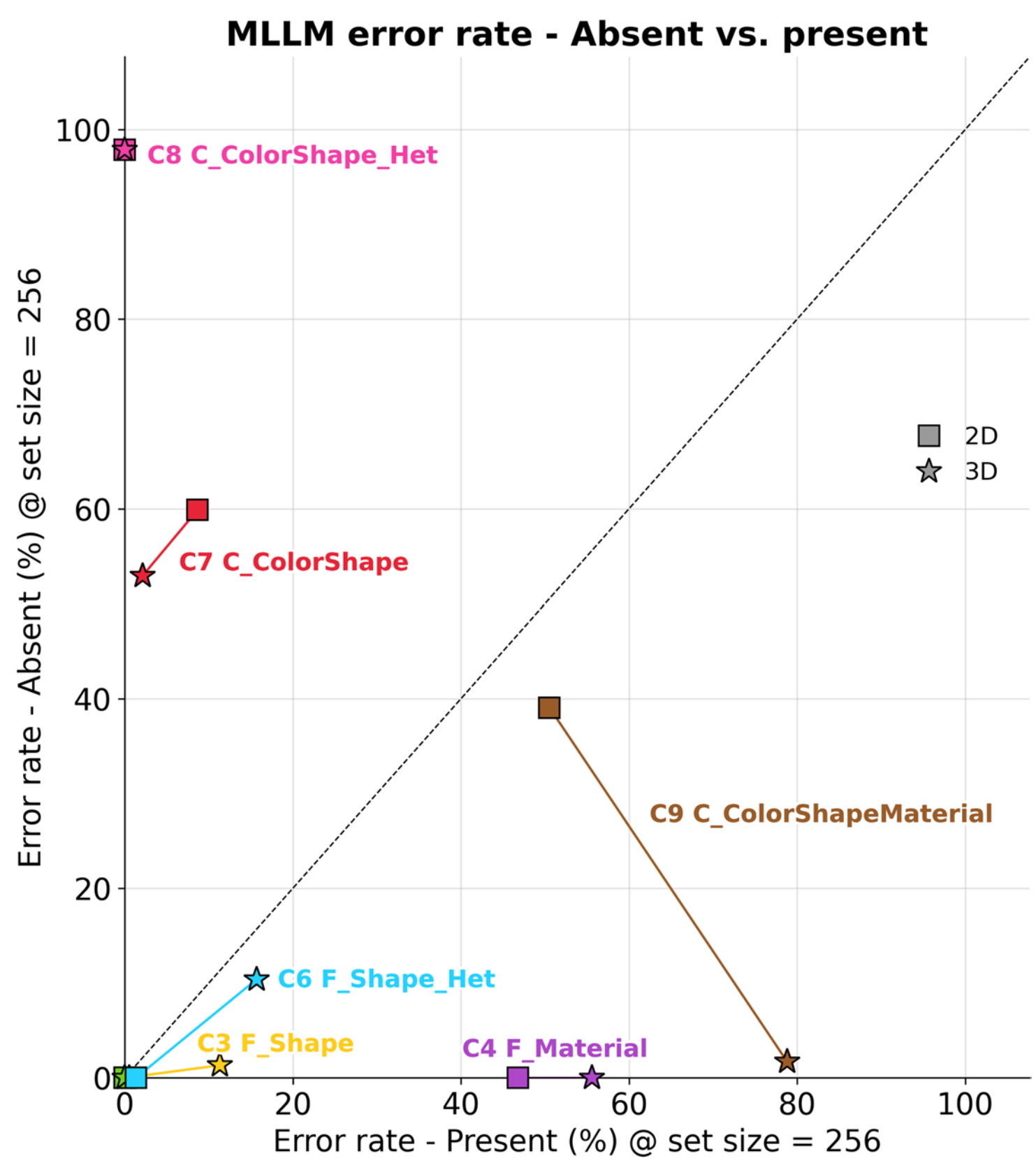

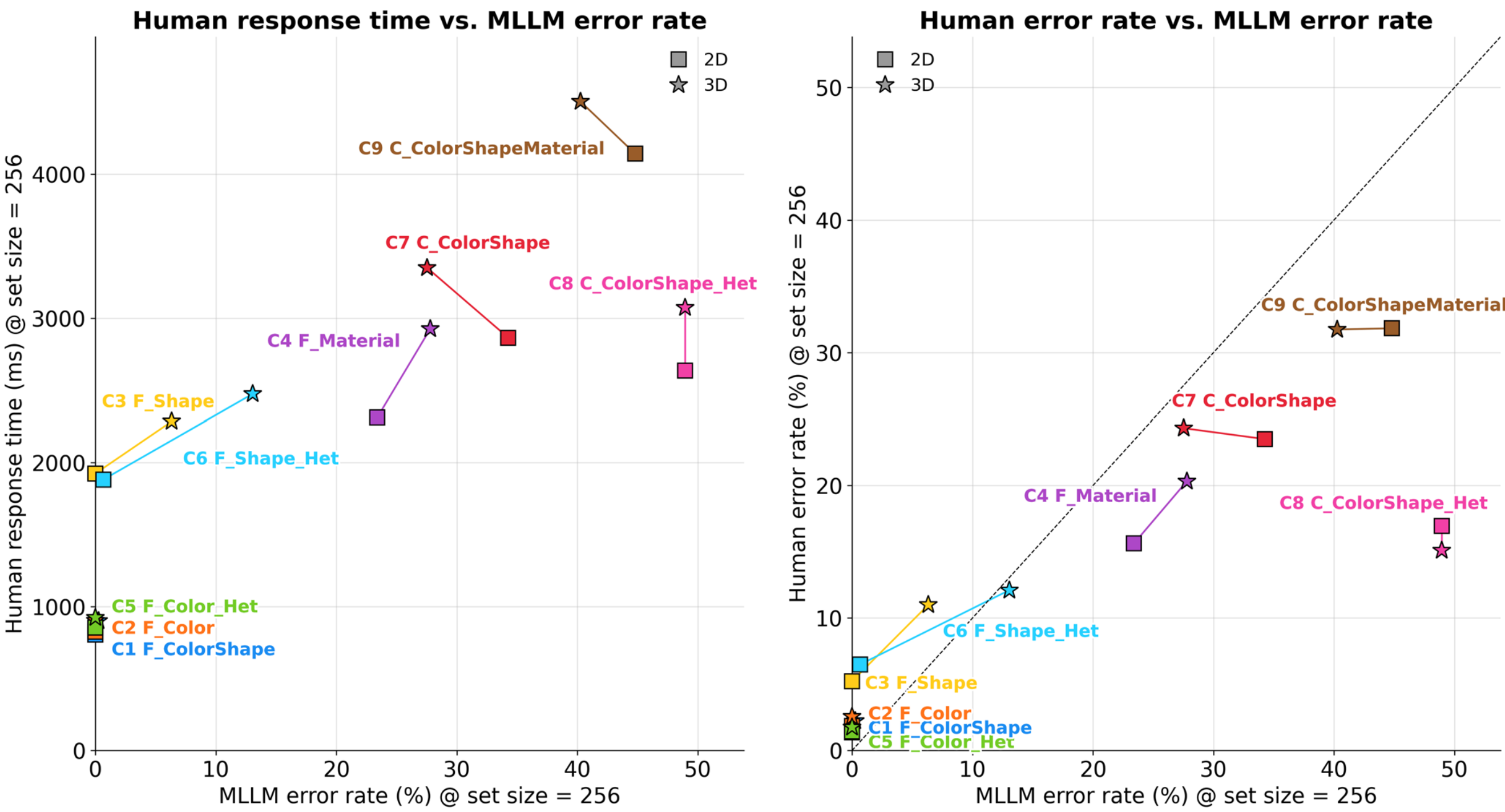

**Fig. 7** Left: Human response time versus MLLM error rate, at the largest set size. Right: Human error rate versus MLLM error rate, at the largest set size. Corresponding 2D and 3D conditions are connected by a line. The diagonal line in the right figure is the line of unity. At this set size, the Spearman correlations were ρ = 0.88 (human RT vs. MLLM error) and ρ = 0.87 (human error vs. MLLM error)

Feature-wise comparisons reinforce this alignment. Color search was most efficient for both groups in 2D and 3D. Shape search was moderately harder, particularly in 3D. Consistent with prior literature (Wolfe & Myers, 2010), material search was the least efficient feature condition for both humans and MLLMs, validating the cross-modal robustness of these visual constraints. Finally, an analysis of performance as a function of target location (Appendix, Fig. 8) revealed a further divergence: human error rates and response times showed a center-of-screen advantage and a left-side disadvantage (for 3D response times, both edges were slowed), whereas MLLM error rates were largely invariant to the target's position in the display.

## Discussion

This study compared visual search in humans and MLLMs using identical synthetic stimuli to determine if models exhibit human-like difficulty signatures. We hypothesized that MLLM error rates would mirror human performance (remaining efficient for feature pop-out but degrading with set size for conjunctions). Our central finding reveals a clear similarity: humans and MLLMs performed well in feature searches but showed systematic degradation in complex conjunction tasks.

Data obtained from 1,250 participants replicated classic findings. Feature searches defined by a unique color (Conditions C1, C2, C5) were efficient, with nearly flat RT slopes, demonstrating the classic pop-out effect (Treisman & Gelade, 1980). In contrast, searches requiring the binding of features (C7–C9) or the discrimination of less salient features such as shape (C3, C6) and material (C4) produced substantial set-size effects on RT, indicative of an effortful capacity-limited process. The addition of 3D depth cues increased search times and steepened slopes, consistent with findings that additional cues such as shadows impair search efficiency (Rensink & Cavanagh, 2004). The 3D cues make search harder by introducing occlusion and clutter, and by limiting the size of the Functional Visual Field (FVF), the area where an item can be successfully identified in a single fixation (Wolfe, 2021).

The performance of the MLLMs paralleled the human patterns with remarkable accuracy. In feature pop-out conditions, MLLM accuracy was high and stable regardless of set size. In conjunction conditions, accuracy decreased as set size increased, with the steepest decreases observed in the most heterogeneous conditions (C8, C9). This suggests that, with current foundation model architectures and pre-training on vast datasets of visual and textual information, MLLMs have implicitly learned the same principles of stimulus similarity and heterogeneity that govern human attentional guidance (Duncan & Humphreys, 1989; Wolfe, 2021). The models' difficulty with conjunctions also echoes recent findings that MLLMs struggle with the binding problem (Alonso et al., 2025; Campbell et al., 2024; Savietto et al., 2026; Tangtartharakul & Storrs, 2026), a core challenge for human attention.

Our findings extend the understanding of search in three-dimensional space. The fact that MLLMs found 3D scenes more challenging suggests their internal representations capture some of the complexity introduced by occlusion, perspective, and shadows, though their failure mode in the C9-3D condition indicates these representations may be fragile. This aligns with findings that MLLMs have deficits in tasks involving occlusion and 3D reasoning (Tangtartharakul & Storrs, 2026).

The results also revealed a divergence between human and MLLM performance, particularly in how they handle the absence or presence of a target. Humans responded more slowly yet more accurately on target-absent trials than on target-present trials (Fig. 4), a pattern consistent with an adaptive quitting threshold (Wolfe, 2021). Humans search until an internal quitting signal reaches this threshold, taking longer on absent trials to be confident the target is not there. In the most difficult conditions, the absent-trial slowing remained below the classic twofold pattern, suggesting that participants lowered this threshold under high load. The MLLMs showed an opposite error-rate pattern in some of the conjunction searches (C7 and C8), with a high error rate for target-absent trials and a low error rate for target-present trials. Conversely, in the most complex condition (C9) in 3D, the models performed near-perfectly on target-absent trials but made many errors when the target was present (up to 79% error rate in the 3D condition). This suggests a breakdown in the models' ability to conduct exhaustive verification of the display; because the models produce a single feed-forward response, they cannot, unlike humans, invest additional search time to secure accuracy on difficult trials. Instead of thoroughly scanning the display, the MLLMs appeared to default to a "present" or "absent" response when faced with visual complexity. A similar present/absent reversal has been reported in reasoning models using token counts as an effort measure (Wick, 2026), suggesting the pattern generalizes across model types and dependent measures. Our spatial analysis (Appendix, Fig. 8) reinforces this divergence: whereas human performance varied with target location in ways consistent with foveated scanning starting at the central fixation cross and with stimulus-response key mapping (a center-of-screen advantage and a left-side disadvantage), MLLM errors were more spatially uniform. This suggests that the models' extreme present/absent biases arise at the response or decision level rather than from position-specific perceptual limits, and that MLLMs lack the embodied constraints (central fixation, saccadic sampling, manual responding) that shape human search.

The strategic mechanism in human search and its absence in the MLLMs is most clearly observed in the performance difference on the C8 condition (a yellow square among yellow circles, blue squares and blue circles). There is a reversal in difficulty: for humans C8 was easier than C7 (Figs. 2 and 3), whereas for MLLMs, it had the highest error rate among all conjunction search conditions (Fig. 5). It is likely that human participants employed a 'subset search' strategy (e.g., Egeth et al., 1984; Friedman-Hill & Wolfe, 1995). Although C8 is more heterogeneous than C7, the addition of blue circles allowed participants to apply a guiding template for yellow, effectively filtering out all blue items (squares and circles). This filtering corresponds to the first stage of the two-template process in Guided Search 6.0 (Wolfe, 2021). In the second stage, a 'Target Template' is used to serially verify the shape of the remaining candidates. This reduced the effective set size in C8 compared to C7, a top-down filtering advantage that the MLLMs do not exploit. In contrast, the MLLMs process the tokenized image and prompt text in a single feed-forward pass, with no opportunity to filter the display and then re-interrogate the remaining candidates.

The graded difficulty we observed across our conjunction conditions supports human perception models that view search efficiency as a continuum rather than a simple serial-parallel dichotomy. Our findings are well-explained by frameworks such as Guided Search (Wolfe, 2021; Wolfe et al., 1989), where distractor heterogeneity reduces the effectiveness of guidance. Specifically, guidance is strongest when featural differences between the target and distractors are large and differences among distractors are small (Duncan & Humphreys, 1989). Our C8 and C9 conditions increased distractor heterogeneity, which should weaken guidance. The MLLM error rates followed this prediction: C8 and C9 were the models' two most difficult conditions. The results also align with the Distractor Rejection Cost model (Xu et al., 2021), which posits that the cost of rejecting distractors is modulated by both target-distractor similarity and distractor-distractor interactions.

Several limitations should be considered. First, our human data were collected online, which introduces variability in viewing conditions and participant engagement compared to a laboratory setting. However, the clear replication of classic visual search effects suggests the data are robust. Second, our evaluation of MLLMs relied on error rate as a proxy for cognitive effort, as RT is not a meaningful measure for these models, although reasoning-token counts may serve as a partial reaction-time analog in reasoning models (Wick, 2026). Third, our stimuli, while more complex than simple letters, were still synthetic. Search in natural, cluttered scenes involves additional guiding factors, such as scene grammar and object co-occurrence, that were not tested here (Wolfe et al., 2011). Finally, our study examined the baseline capabilities of MLLMs. We did not employ iterative prompting, visual scaffolding techniques, or reasoning MLLMs that have been shown to improve MLLM performance on search-like tasks (Izadi et al., 2025; Wu & Xie, 2024). In addition, our tests were limited to three models, all produced by the same company. The behavior of MLLMs may change over time and models developed by other authors may perform differently.

In conclusion, our comparative study demonstrates that MLLMs have implicitly learned many of the fundamental principles that govern human visual search. Their performance is affected by target-distractor similarity, distractor heterogeneity, and the feature-conjunction distinction in a manner that closely mirrors human RT data. This suggests that the statistical patterns in large-scale visual-language datasets are sufficient to instill human-like sensitivities to the factors that guide attention. However, a difference appears in the most complex search scenarios, particularly in target-absent decisions, revealing a potential weakness in the models' ability to handle combined feature binding and high distractor load.

One promising avenue for further research is to compare human fixation patterns directly with the internal attention maps of MLLMs using the same stimuli to see if the models' "attention" corresponds to human gaze (Christian et al., 2026). Further investigation is also needed to understand the MLLMs' failure in complex target-absent conditions. Is this a failure of binding, a premature quitting strategy, or an artifact of the models' training? Additionally, extending the current comparative paradigm to include other known modulators of human search, such as scene context, search history, and reward (Wolfe & Horowitz, 2017), will be important for developing a more complete understanding of the similarities and differences between biological and artificial visual intelligence.

## Appendix A. Construction of Images

We developed a configurable generator with parallel branches for 2D images and 3D scenes (based on the CLEVR framework; Johnson et al., 2017) rendered in Blender (Blender Foundation, 2017). Stimuli were constructed from three features combined according to condition:

**Shape & Size** 2D stimuli used circles (radius 28px) and squares (side 49px). 3D stimuli used spheres and cubes, calibrated so the projected pixel area of a central sphere matched the 2D circle. To account for perspective, 3D object sizes varied with distance.

**Color** Shapes were bright yellow (RGB [255, 255, 0]) or blue. In 2D, a dark blue (RGB [0, 0, 180]) was used; in 3D, a brighter blue (RGB [30, 30, 255]) was selected to compensate for shading and reflection.

**Material** Objects were either solid or textured. The texture (Gaussian noise applied to 1/6 of the surface) was identical in 2D and 3D (custom Blender material).

Images were generated at 1920 × 1080 resolution. Objects were placed sequentially (target first) within a central display area, enforcing minimum distances to prevent overlap (2D) or collision (3D).

**2D** Shapes were placed upright without rotation.

**3D** The camera was positioned at 33.7° to reveal front, top, and side faces, balancing visibility with depth perception. Three-point lighting ensured distinct shadows. Partial occlusion was permitted as a natural depth cue.

We generated 25 versions per configuration (condition × target presence × set size × dimensionality). Distractor types were balanced; if counts were not perfectly divisible, remaining slots were filled sequentially. Target locations (and 3D pixel coordinates) were recorded for analysis.

## Appendix B. Task Instruction

**Task instruction for 2D images**:

In this task, you will search for a **solid yellow square** (pictured below) in a series of images.

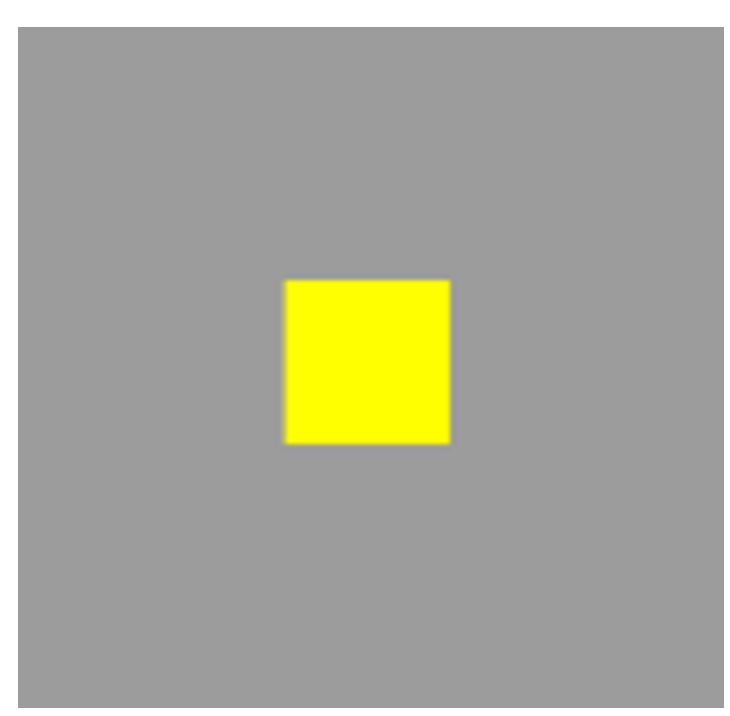

The experiment will begin with a practice block of 9 images, followed by the main experiment, which consists of 9 blocks of 14 images each. You can take a short break between blocks.

The sequence for each trial is as follows:

- A fixation cross (+) will appear in the center of the screen. Please keep your eyes focused on this cross.
- An image will then be displayed. Your task is to decide if the solid yellow square is **absent** or **present**.
- Respond as **quickly and accurately as possible** by pressing the correct key:
  - Press the **A** key if the object is **Absent**.
  - Press the **P** key if the object is **Present**.
- After your response, you will receive feedback: the word 'Correct' will appear in green letters, or the word 'Incorrect' will appear in red letters.

Press the **Space** key to begin the practice block.

**Above each 2D image**:

Solid yellow square: press **A** for **absent** or **P** for **present**.

**Task instruction for 3D images**:

In this task, you will search for a **solid yellow cube** (pictured below) in a series of images.

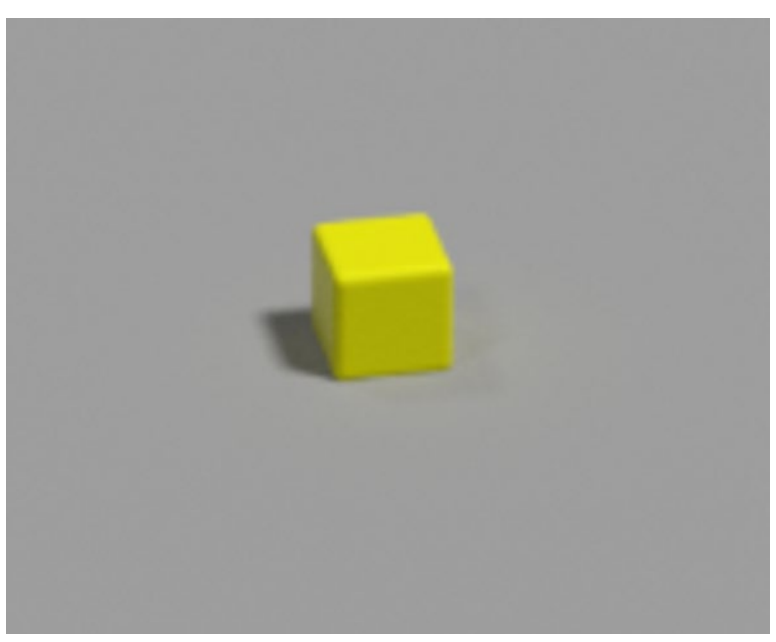

The experiment will begin with a practice block of 9 images, followed by the main experiment, which consists of 9 blocks of 14 images each. You can take a short break between blocks.

The sequence for each trial is as follows:

- A fixation cross (+) will appear in the center of the screen. Please keep your eyes focused on this cross.
- An image will then be displayed. Your task is to decide if the solid yellow cube is **absent** or **present**.
- Respond as **quickly and accurately as possible** by pressing the correct key:
  - Press the **A** key if the object is **Absent**.
  - Press the **P** key if the object is **Present**.
- After your response, you will receive feedback: the word 'Correct' will appear in green letters, or the word 'Incorrect' will appear in red letters.

Press the **Space** key to begin the practice block.

**Above each 3D image**:

Solid yellow cube: press **A** for **absent** or **P** for **present**.

## Appendix C. Prompt for MLLM Evaluation

To mirror human instructions while resolving model confusion between solid and textured targets observed in pilots, we explicitly clarified the textures. For each model, each image was queried 10 times at temperature 0:

In this task, you will search for a solid yellow {block} without any texture on its face in an image.

{variations}

Solid yellow {block}: Output 0 for absent or 1 for present.

The placeholder {block} was adapted to the dimension (square/cube), and {variations} consisted of one to ten appended dots, providing ten deterministic prompt variants.

## Appendix D. Spatial Distribution of Target Locations

We analyzed whether target location influenced difficulty. For every target-present image we computed the per-image error rate (for humans and MLLMs) and, for humans, the mean response time, and standardized each metric within its set (9 conditions × 7 set sizes × 2 dimensionalities) of 25 images as a $z$-score. Sets without variance were assigned $z = 0$. This applied to 71 of the 126 MLLM sets, in all of which the models made no errors on any of the 25 images, and to none of the human sets. Each image was then plotted at its target's on-screen pixel location and colored by this $z$-score (green = below-average, grey = average, red = above-average error or response time; Fig. 8).

**Humans** Both error rate and response time (Fig. 8, top two rows) showed a modest center-of-screen advantage and a left-side bias, with left-side targets being harder and slower. Relative to central targets, right-edge targets were also slower (in both dimensionalities) and, in 3D, more error-prone; the left-right asymmetry was thus strongest for 2D error rates and absent for 3D response times.

**MLLMs** In contrast, MLLM error rates remained close to the condition average across the display, indicating that performance is largely invariant to spatial position and lacks the biological or motor biases observed in humans (Fig. 8, bottom row).

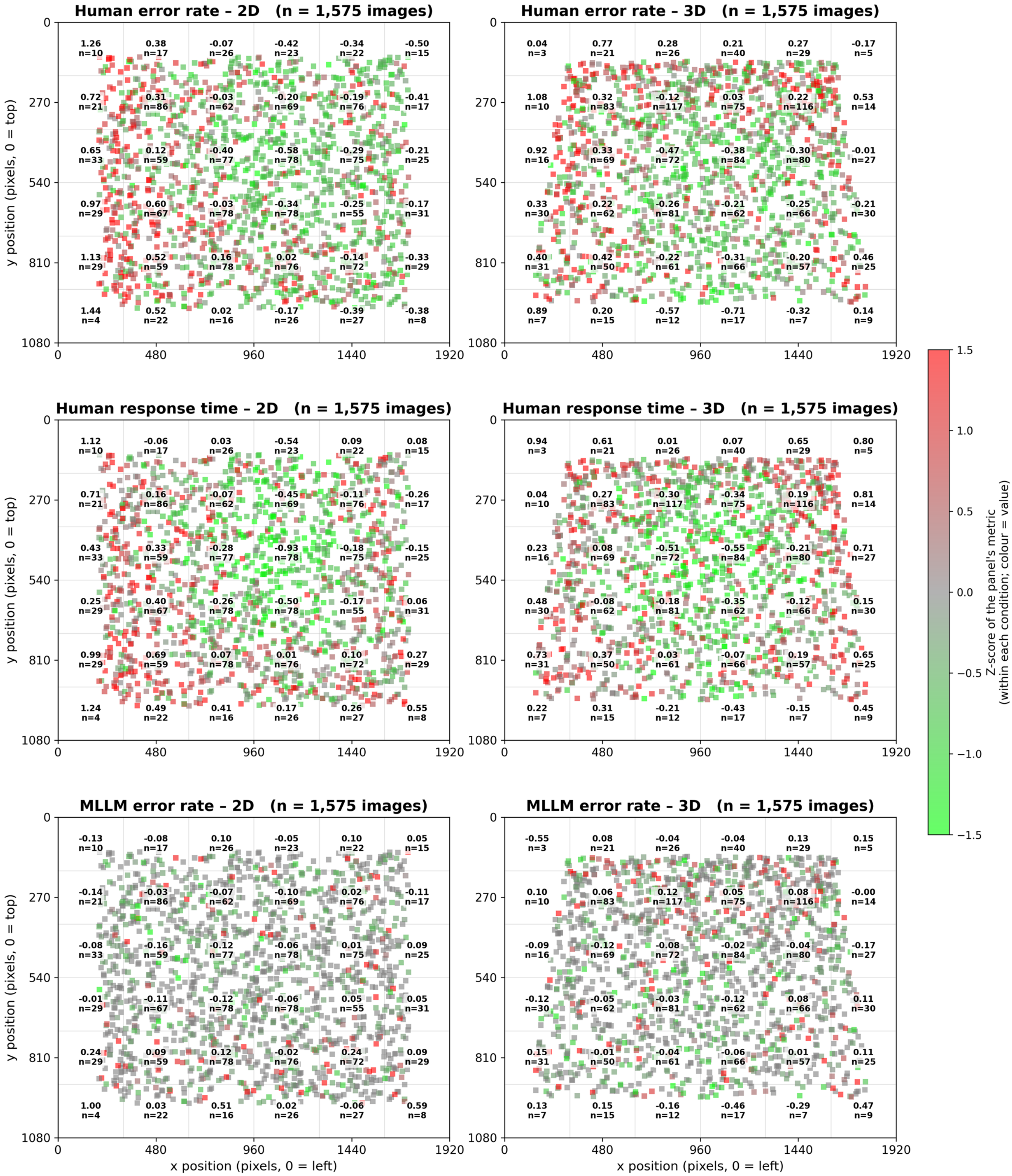

**Fig. 8** Spatial distribution of target locations, shown separately for human error rate, human response time, and MLLM error rate (rows) in 2D and 3D (columns). Each square marks one target-present image at its on-screen pixel location, colored by the per-condition *z*-score of that metric (green = below-average, grey = average, red = above-average; sets without variance were assigned z = 0). Overlaid text reports the mean *z*-score and the number of images (*n*) in each spatial bin

**Author contributions** R.Z. and J.d.W. co-designed the study and analyzed the data. R.Z. created the stimuli and tested the MLLMs. J.d.W., H.S., D.D., and Y.B.E. provided supervision and facilitated critical discussion. All authors reviewed the manuscript.

**Funding** This project has received funding from a Cohesion Project of the Faculty of Mechanical Engineering, Delft University of Technology.

**Data Availability** The data and analysis code underlying this work are available at https://doi.org/10.4121/e2b3e80f-8983-478b-a93f-8c8858fe8f96.

## Declarations

**Ethics Approval** This study received ethical approval from the TU Delft Human Research Ethics Committee (Reference Number: 4182).

**Consent to Participate** Informed consent was obtained from all respondents through a dedicated item at the beginning of the questionnaire.

**Competing Interests** The authors declare no competing interests.